\documentclass[mathleft
]{an}
\usepackage{graphicx}
\usepackage{times}
\begin{document}

% The following seven commands are intended for editorial usage and should be ignored by
% the author(s).
\Pagespan{789}{}% Document's page range. 
% If second parameter is left empty, the last page is computed automatically.
\Yearpublication{2015}%
\Yearsubmission{2015}%
\Month{09}%   
\Volume{999}%  
\Issue{88}% 
\DOI{This.is/not.aDOI}% 

\title{A New Characterization of the Compton Process in the ULX Spectra}

\author{S. Kobayashi\inst{1}\fnmsep\thanks{Corresponding author:
  \email{kobayashi@juno.phys.s.u-tokyo.ac.jp}\newline}
%Example 
%for footnote, note the usage of the \texttt{fnmsep}
%command as separator between institute number and footnote mark} 
\and K. Nakazawa\inst{1} \and K. Makishima\inst{2}
}
\titlerunning{Instructions for authors}
\authorrunning{S. Kobayashi \and K. Nakazawa \and K. Makishima}
\institute{
Department of Physics, The University of Tokyo, 7-3-1 Hongo, Bunkyo-ku, Tokyo-to 113-0033, Japan
\and 
MAXI team, RIKEN, 2-1 Hirosawa, Wakou-shi, Saitama-ken 351-0198, Japan
}

\received{1 Nov 2015}
%\accepted{11 Nov 2015}
\publonline{later}


\abstract{
Attempts were made to construct a unified description of the spectra of ULX (Ultra Luminous X-ray source) objects, including their Power-Law (PL) state and Disk-like state. Among spectral models proposed to explain either state, the present work adopts the one which combines multi-color disk (MCD) emission and its thermal Comptonization (THC). This model was applied to several datasets of ULXs obtained by {\it Suzaku}, {\it XMM\--Newton}, and {\it Nustar}. The model well explains all the spectra, regardless of the spectral states, in terms of a cool disk ($T_{\rm{in}}=0.2\--0.5$ keV) and a cool thick ($T_{\rm{e}}=1\--3$ keV, $\tau \sim10$) corona. The fit results can be characterized by two new parameters. One is $Q\equiv T_{\rm{e}}/T_{\rm{in}}$ which describes balance between the Compton cooling and gravitational heating of the coronal electrons, while the other is $F\equiv 1-F_{\rm{direct}}/F_{\rm{total}}$, namely, the covering fraction of the MCD by the corona. Here, $F_{\rm{direct}}$ and $F_{\rm{total}}$ are luminosity in the directly-visible disk emission and the total radiation, respectively. Then, the PL-state spectra have been found to show $Q\sim10$ and $F\sim0.5$, while those of the Disk-like state $Q\sim 3$  and  $F\sim1$. Thus, the two states are clearly separated in terms of $Q$ and $F$. The obtained results are employed to argue for their interpretation in terms of high-mass (several tens to several hundreds $M_{\rm{\odot}}$) black holes.
}

\maketitle

\section{Introduction}
Ultra Luminous X-ray sources (ULXs) are unusually luminous X-ray objects located at off-center regions of spiral galaxies. Their luminosity, $L_{X}=10^{39.5-41}$ erg sec$^{-1}$, exceeds the Eddington limit $L_{\rm{Edd}}$ of a $10~M_{\rm{\odot}}$ black hole (BH) by $1\--2$ orders of magnitude. Although they are thought to be accreting BHs, their masses are still controversial since their discovery (Fabbiano~1989).
While their high luminosity can be interpreted as sub-Eddington emission from relatively massive ($100\--1000~M_{\rm{\odot}}$) BHs (Makishima et al.~2000), it can alternatively be explained as super-Eddington emission from ordinary stellar mass ($\sim10~M_{\rm{\odot}}$) BHs (e.g., Mineshige \& Ohsuga~2007). The most effective way to settle the controversy is to estimate the masses of these BHs by detecting orbital doppler shifts in their optical counterparts. Although such attempts are providing some initial results (e.g., Motch et al.~2014), these measurements are still limited in number. Therefore, X-ray spectroscopy continues to play an important roll in estimating the ULX mass.

Like BH binaries (BHBs), ULXs show two different spectral states depending on their luminosity. One is called Disk-like state characterized by spectra of round convex shapes. The other is called Power-Law (PL) state, in which the spectra literally show PL shapes with a mild cut-off at around 8 keV; in addition, occasionally a soft excess appears at $\sim 1.5$ keV which makes the spectra \textquotedblleft two-humped\textquotedblright . Some objects make transitions between the two states (e.g., Kubota, Makishima \&Ebisawa~2001), and are known to be dimmer in the PL-state. In detail, the PL-state may be further classified into two sub-states based on the PL shape (Sutton, Roberts \& Middleton~2013). However, we treat them together since the sample studied in this work is small.

So far, the PL-state with the \textquotedblleft two-hump\textquotedblright \ property has been explained with several alternative ways. While, Middleton et al.~(2015a) and Kawashima et al.~(2012) propose that the soft excess is originates from optically thick winds which are launched from a hot accretion disk due to super-critical accretion, some other authors including Gladstone, Roberts \& Done~(2009), and Feng \& Soria~(2011), try to explain such spectra in terms of a combination of a Multi-Color Disk (MCD) model and its thermal Comptonization (THC), like other BHBs. Although several ULX spectra in the PL-state can be reproduced with the former modeling (Walton et al.~2014,~2015, Bachetti et al.~2013), we have adopt the latter standpoint, because ULX spectra are generally too featureless (at most very weak features; Middleton et al.~2013, Middleton et al.~2015b) and too stable at low energies to be consistent with the former view. 

The Disk-like state has often been explained with a modified MCD model, so called Slim disk model (Watarai et al.~2000). This is an accretion mode thought to emerge at $L_{\rm{X}}/L_{\rm{Edd}} \ge 1$. However, Miyawaki et al. (2009) pointed out that this Slim disk interpretation requires extreme conditions, including too high a disk temperature in the case of M82 X-1, one of the most luminous ULXs. In addition, they found that the MCD+THC model instead derives more physically appropriate parameters from M82 X-1, which exhibited a typical Disk-like state spectra during {\it{Suzaku}} observations. In the present work, we would like to examine whether the view of Miyawaki et al. (2009) is still valid among other ULXs. For this purpose, we apply the MCD+THC model to several ULX spectra in the Disk-like state, as well as to those in the PL-state. 

\section{Targets and data reduction}
As tabulated in Table\ref{datasets}, we selected data sets of several ULXs from nearby galaxies, and analyzed them. The data were derived from {\it Suzaku}, {\it XMM\--Newton}, and {\it Nustar} archives, and were all acquired in normal observational modes of the respective observatories. 
\begin{table*}[ht!]
\centering%%%
\caption{The list of data sets used in the present study.}
\label{datasets}
\begin{tabular}{cccccc}\hline
Target & Satellite & Obs.Date & Exposure (ks) & ObsID & ID\\
\hline
Hol IX X-1 & {\it XMM\--Newton} & 2001/04/23 & 83 & 0111800101 & A\\
		    & {\it Suzaku} & 2012/04/13 & 180 & 707019010 & B\\
		    & {\it Suzaku} & 2012/10/24 &  $183+217$ & 707019030 \& 707019040 & C\\
\hline
Hol II X-1  & {\it XMM\--Newton}+{\it Nustar} & 2013/09/17 & 14 \& 111 & 0724810301 \& 30001031005 & D\\
		    & {\it XMM\--Newton} & 2004/04/15 & 82 & 0200470101 & E\\
\hline
NGC1313 X-1 & {\it Suzaku} & 2008/12/05 & 91 & 703010010 & F\\
		 	 & {\it XMM\--Newton}+{\it Nustar} &  2012/12/16 & 125 \& 101 & 0693850501 \& 30002035002 & G\\
			 & {\it XMM\--Newton}+{\it Nutar} & 2012/12/22 & 125 \& 127 & 0693851201 \& 30002035004 & H\\
\hline
M33 X-8	    & {\it Suzaku} & 2010/01/11 & 106 & 704016010 & I\\
\hline
\end{tabular}
\end{table*}
We used the entire exposure in each data set to maximize the statistics. The target spectra were accumulated inside circle regions of which the radii were chosen to be slightly larger than the PSF of the respective telescopes (e.g. $1'.5$ for {\it Suzaku}). The background spectra were accumulated from source-free regions near the target. In the {\it Suzaku} XIS data analysis, we excluded the $1.7\-- 2.0$ keV energy band to avoid instrumental structures due to Si edge. The spectral analysis was done with XSPEC ver. 12.8.2.

\subsection{Holmberg IX X-1}
Holmberg (Hol) IX is a dwarf galaxy at a distance of 3.4 Mpc (Georgiev et al.~1991), associated with the spiral galaxy M81, and X-1 is a ULX located in it. It is one of the brightest ULXs in nearby galaxies, with an average luminosity of $L_{\rm{X}} \sim 10^{40}$ erg sec$^{-1}$. Thanks to deep observations by many observatories, it is known to show high variability on time scales of month to years (Walton et al~2014). State transitions were confirmed among observations (La Parola et al. 2004; Tsunoda et al.~2006; Vierdayanti et al.~2009).
\subsection{NGC1313 X-1}
This is a ULX associated with the spiral galaxy NGC 1313 (4.13 Mpc; M\'endez et al.~2002). It is also known to show high variability. Its luminosity reaches $L_{\rm{X}}=3.5\times10^{40}$ erg sec$^{-1}$ at maximum, while the average is $L_{\rm{X}}\sim3\times10^{39}$ (Mizuno et al.~2007). There is another ULX called X-2 at $\sim6'$ off X-1. 
\subsection{Holmberg II X-1}
It is a ULX associated with the dwarf galaxy Hol II (3.39 Mpc; Karachentsev et al.~2002). While the ULXs described above show hard spectra ($\Gamma=1.6\--2.0$) in the PL-state, this ULX shows relatively soft spectra ($\Gamma=2.0\--2.4$) with strong soft excess. These are likely to belong to the \textquotedblleft soft ultraluminous state\textquotedblright \ which is introduced by Sutton et al. (2013).
\subsection{M33 X-8}
This is a ULX located in a slightly off-center position of the spiral galaxy M33. Because it is so close to the galaxy center, it was at first considered to be an Active Galactic Nucleus (AGN). However, thanks to $\it{ASCA}$, its spectrum was found to have a convex shape, and is different from those of AGNs (Takano et al.~1994). It has hence been regarded as the closest (795 kpc; van den Bergh 1991) and one of the dimmest ($L_{\rm{X}}\sim10^{39}$ erg sec$^{-1}$ in average) ULXs that have been studied in detail (Makishima et al. 2000). This object is expected to connect ULXs and the ordinary BHBs because of its low luminosity.

\section{Results}
\begin{figure*}
\begin{center}
\includegraphics[width=150mm,height=200mm]{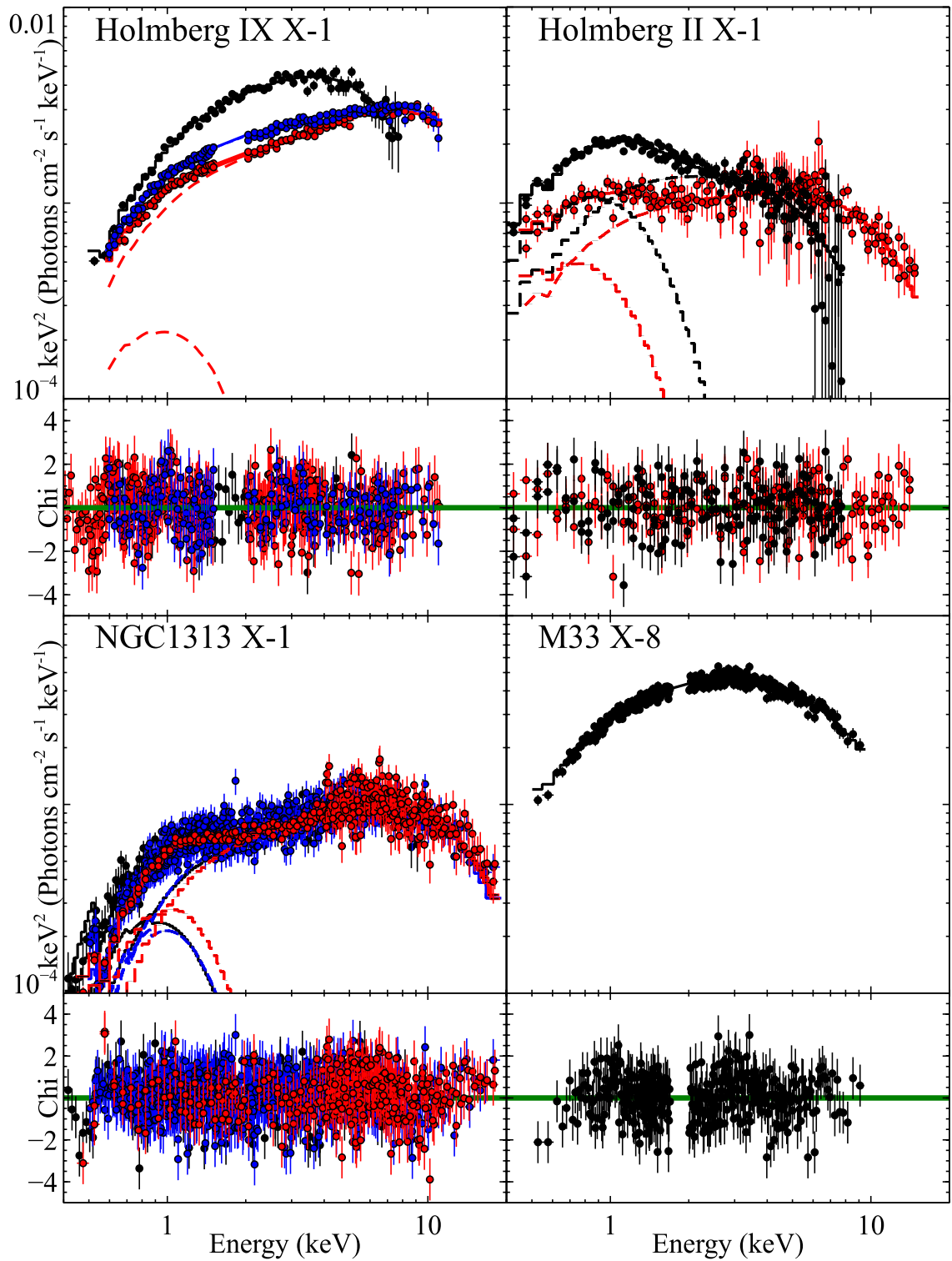}
\caption{Spectra of the ULXs studied here, in $\rm{\nu F \nu}$ form, each fitted with the MCD+THC model. Bottom panels show residuals from the model in each spectrum. Different observations of the same object are specified by colors in the time order of black, red, and blue.}
\label{label1}
\end{center}
\end{figure*}

\begin{table*}[ht]
\centering%%%
\caption{Summary of fitting using the MCD+THC; $\tt{phabs}(\tt{diskbb}+\tt{nthcomp})$ model. The errors refer to $90\%$ confidence level.}
\label{fitresults}
\begin{tabular}{@{~}c@{~}c@{~~}c@{~~}ccccccccc}\hline
\hline
Target & ID$^{a}$ & State & $N_{\rm{H}}^{b}$ & $T_{\rm{in}}$ (keV) & $T_{\rm{e}}$ (keV) & $T_{\rm{e}}/T_{\rm{in}}$ & $\Gamma^{c}$ & $\tau^{d}$ & $L_{\rm{X}}^{e}$ & $F$ & $\chi^{2}/\nu$\\
\hline
Hol IX X-1 & B & PL & $0.97^{+0.03}_{-0.02}$ & $0.28\pm0.02$ & $2.7\pm0.2$ & $10\pm1$ & $1.66\pm0.02$ & $\sim14$ & 18.4 & 0.62 & 501/408\\
		    & C & PL & $1.29^{+0.05}_{-0.04}$ & $0.25^{+0.05}_{-0.03}$ & $2.9\pm0.2$ & $11\pm2$ & $1.76\pm0.01$ & $\sim12$ & 21.8 & 0.82 & 484/408\\
		    & A & Disk & $1.40\pm0.06$ & $0.31\pm0.02$ & $1.34^{+0.13}_{-0.08}$ & $4.3^{+0.5}_{-0.4}$ & $1.70\pm0.01$ & $\sim20$ & 22.5 & 1.0 & 81.4/69\\
\hline
Hol II X-1  & D & PL & $0.8^{+0.7}_{-0.5}$ & $0.21^{+0.06}_{-0.04}$ & $2.6\pm0.2$ & $12\pm4$ & $1.92^{+0.05}_{-0.06}$ & $\sim11$ & 8.4 & 0.44 & 218/200\\
		    & E & Disk & $0.79\pm0.03$ & $0.294^{+0.015}_{-0.003}$ & $1.2\pm0.2$ & $4.1\pm0.7$ & $1.9\pm0.1$ & $\sim17$ & 9.5 & 0.77 & 843/786\\
\hline
NGC1313 X-1 & G & PL & $2.6\pm0.3$ & $0.21^{+0.03}_{-0.02}$ & $2.9\pm0.2$ & $14\pm2$ & $1.79\pm0.05$ & $\sim12$ & 8.7 & 0.31 & 382/322\\
 			  & H & PL & $2.4\pm0.4$ & $0.23\pm0.03$ & $2.8\pm0.2$ & $12\pm2$ & $1.81\pm0.04$ & $\sim12$ & $9.2$ & 0.38 & 625/679\\
			  & F & PL &  $1.7^{+0.6}_{-0.5}$ & $0.23\pm0.04$ & $2.1^{+0.3}_{-0.2}$ & $9.4\pm1.4$ & $1.68\pm0.06$ & $\sim15$ & $8.4$ & 0.35 & 240/196\\
\hline
M33 X-8         & I & Disk & $0.4\pm0.1$ & $0.55\pm0.06$ & $2.0\pm0.3$ & $3.6\pm0.4$ & $2.1\pm0.1$ & $\sim12$ & $1.7$ & $ 1.0$ & 357/330\\
\hline
\end{tabular}
\begin{flushleft}
a: Data set ID which is defined in table\ref{datasets}.\\
b: Equivalent hydrogen column density of the absorber, in units of $10^{21}$ atoms cm$^{-2}$.\\
c: Photon index of the Comptonization component.\\
d: Optical thickness of the Compton cloud, derived with equation (2). \\
e: The absorption corrected $0.1\-- 10$ keV luminosity, in units of $10^{39}$ erg sec$^{-1}$.
\end{flushleft}
\end{table*}

Figure \ref{label1} shows the spectra of all data sets, sorted according to the targets. Out of the 9 data sets of the 4 objects, three can be readily identified with the Disk-like state; one (black) from Hol IX X-1, another from Hol II X-1 (black), and the other from M33 X-8. The other 6 spectra can be regarded as in the PL-state. While the three Disk-like state spectra exhibit round shapes, those assigned to the  PL-state show flatter continua with turn-over at $\sim 8$ keV, and some soft excess is seen in $0.8\-- 1.5$ keV. All spectra in the Disk-like state have higher luminosity than those in the PL-state of the same object. None of the spectra show clear evidence for line or absorption. 

We fitted all spectra with the MCD+THC model, expressed as $\tt{phabs}*(\tt{diskbb}+\tt{nthcomp})$ in XSPEC. Here the factor $\tt{phabs}$ represents photoelectric absorption which cross-sections by Baluci\'nska-Church \& McCammon~(1992) and Yan, Sadeghpour \& Dalgarno~(1998). The other two components, $\tt{diskbb}$ and $\tt{nthcomp}$ (Zdziarski, Johnson \& Magdziarz~1996, Zycki, Done \& Smith~1999), describe multi-color blackbody radiation from a standard accretion disk, and THC of the photons supplied from the disk, respectively. Thus, we are considering a condition wherein the disk is partially covered with a Compton cloud, and photons from the remaining fraction are coming out un-scattered to form the soft excess at $< 1.5$ keV. The results of the fitting are summarized in table \ref{fitresults}, where errors refer to the $90\%$ confidence level of individual parameters. 

The MCD+THC model was successful in all the spectra regardless of their spectral state, and the parameters of the model were constrained well. (Although there are still some residuals in low energy band, especially around O K-edge at 0.54 keV, these are within uncertainties of instrumental calibration.) The three objects (except M33 X-8) was all brighter by $\sim 10 \%$ in the Disk-like state than in the PL-state. The photon index $\Gamma$ and the coronal electron temperature $T_{\rm{e}}$ are mainly determined by the slope and the high-energy cutoff of the spectrum, respectively. As a result, the round shaped spectra in the Disk-like state were reproduced by lower values of $T_{\rm{e}}\sim 1.6$ keV and slightly steeper slopes as $\Gamma=1.7\--2.1$ than those in the PL-state ($\Gamma=1.6\-- 1.9$, $T_{\rm{e}}\sim 2$ keV). On the other hand, the inner-radius temperature $T_{\rm{in}}$ is constrained by the peak energy of the soft excess, or the curvature in $<1$ keV. This made $T_{\rm{in}}$ closer to $T_{\rm{e}}$ in the Disk-like state to explain the convex shape. As a result, they gave higher temperatures as $T_{\rm{in}}\sim 0.5$ keV than those in the PL-state ($T_{\rm{in}}\sim 0.2$ keV). 

Once we obtain $\Gamma$ and $T_{\rm{e}}$, the optical depth of the corona can be calculated as (Sunyaev \& Titarchuk ~1980)
\[
\tau = \sqrt{2.25+\frac{3}{(T_{\rm{e}}/511\ \rm{keV})[(\Gamma + 0.5 )^{2}-2.25]}}-\frac{3}{2} \ . 
\]
According to this equation, the analyzed ULX spectra all require optically thick ($\tau>10$) and cool (several keV) coronae in common, without much depending on the spectral state. These large values of $\tau$ result from relatively flat $\Gamma$ and very low $T_{\rm{e}}$.

\section{Discussion}
\subsection{Characterization of the Spectral States}
As shown in \S3, we were able to interpret all the nine ULX spectra with a unified MCD+THC model. The obtained model parameters are actually very similar between the two states. Nevertheless the spectral shape clearly depends on the state (Fig. \ref{label1}). Therefore we would like to invent a way to better characterize and differentiate the two states. For this purpose, we here introduce two new parameters.

One is the temperature ratio $Q$ defined as
\begin{equation}
Q \equiv \frac{T_{\rm{e}}}{T_{\rm{in}}}.
\end{equation}
This $Q$ reflects a balance between Compton cooling and ionic heating of the coronal electrons, in other words, the strength of THC. For example, $Q$ becomes larger ($T_{\rm{e}} \gg T_{\rm{in}}$) when seed photons are up-scattered to higher energies, to result in a more efficient THC process. In contrast, if $Q$ becomes closer to 1 ($T_{\rm{e}}\ge T_{\rm{in}}$), the THC process will no longer take place.

The other new parameter is covering fraction $F$ of the disk by the corona, which is expressed as 
\[
F=1-\frac{F_{\rm{direct}}}{F_{\rm{total}}}, 
\] 
where $F_{\rm{direct}}$ and $F_{\rm{total}}$ are fluxes of the directly-visible disk and the total disk, respectively. In the calculation of $F_{\rm{total}}$, we assume that the photon number is conserved in the THC in the coronae. This $F$ represents the fraction of the accretion disk area that is covered with the Compton cloud. 

\begin{figure}
\begin{center}
\includegraphics[width=80mm,height=80mm]{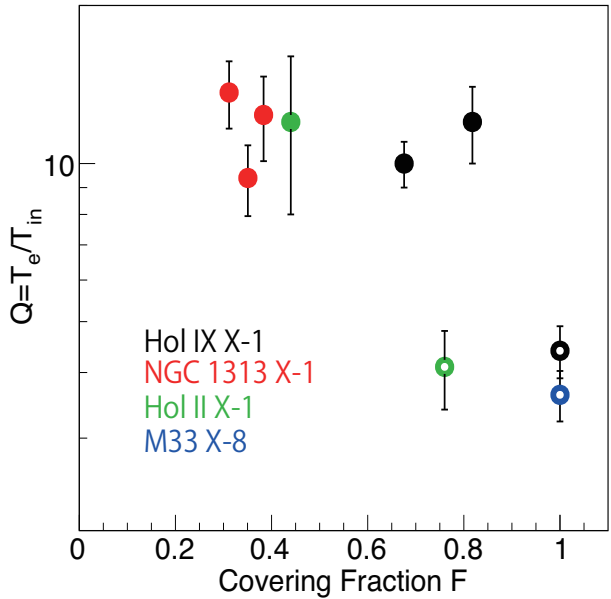}
\caption{The nine spectra characterized on the $Q$ vs $F$ plane. Colors specify the objects. Black: Hol IX X-1. Red: NGC 1313 X-1. Green: Hol II X-1. Blue: M33 X-8. Open circles are the Disk-like state and the filled circles are the PL-state.}
\label{label2}
\end{center}
\end{figure}
In Fig. \ref{label2}, we plot the 9 spectra on the $(Q, F)$ plane to examine how the two new parameters characterize the spectral shape. While the PL-states are located at $(Q, F)=(10, 0.3\--0.8)$, the Disk-like states are at $(Q, F)\sim(3, > 0.7)$. Thus, we succeeded to separate the two states in terms of $Q$ and $F$. The larger values of $Q$ in the PL-state result from larger separations in the spectra between the seed-photon peak ($\propto T_{\rm{in}}$) and the high-energy spectral cutoff ($\propto T_{\rm{e}}$). As the source makes a transition to the Disk-like state, $F$ becomes closer to $1$, because the disk gets fully covered with Compton cloud, and the soft excess disappears. 

Although we were able to separate the two states of the same object, the data points of different objects are still scattered horizontally (along $F$) in Fig. \ref{label2}. This comes from the differences in the strength of the spectral soft excess (the directly-visible disk component). For example, Hol IX X-1 has weaker directly-visible component in the spectrum (Fig.\ref{label1}), so that its data points are shifted systematically to the right (larger $F$). The Disk-like state spectrum of Hol II X-1 has a relatively small $F$, because it exhibits a strong soft excess even in the Disk-like state. 

\subsection{Mass Estimation}

We would also like to estimate the masses of the BHs embedded in these ULXs. Below, we challenge this issue based on two independent arguments. 
\subsubsection{Critical Luminosity of Spectral State Transition}
From table \ref{fitresults}, we can estimate the critical luminosity at which the spectral transition takes place. For example, in Hol IX X-1 it should be somewhere between $2.15\times 10^{40}$ and $2.25 \times 10^{40}$ erg s$^{-1}$, and in M33 X-8 it should be $< 1.7\times 10^{39}$ erg s$^{-1}$, i.e. 10 times smaller. Thus the critical luminosity scatters among the sample ULXs by an order of magnitude. If we assume that the spectral transition takes place at particular Eddington ratio $\eta = L_{\rm{X}}/L_{\rm{Edd}}$ like BHBs, the masses of ULXs should also scatter over a similar range. In other words, ULXs are considered to follow a considerably broad mass spectrum than BHBs.

Motch et al.~(2014) discovered that the low luminosity ($\sim 2\times 10^{39}$ erg sec$^{-1}$) ULX, P13 in NGC 7793, is a $15~M_{\rm{\odot}}$ BH shinning at $\eta\sim 2$. Considering this result, let us assume that the minimum mass of ULXs is $10~M_{\rm{\odot}}$. For example, M33 X-8 could be such an object shining at $\eta \sim 1$. Then, the most massive one in our sample, Hol IX X-1, will reach $> 100~M_{\rm{\odot}}$. Therefore, it is difficult to consider that all the ULXs analyzed have are ordinary stellar-mass BHs. The same argument was already employed by Mizuno et al. (2007). It does not assume whether ULXs are super-Eddington objects or not.

Importantly, the above argument does not depend on the choice of the spectral modeling. It is not affected, either, whether ULXs are super-Eddington objects or sub-Eddington BHs.

\subsubsection{Via Accretion Disk Parameters}
\begin{figure}
\begin{center}
\includegraphics[width=80mm,height=80mm]{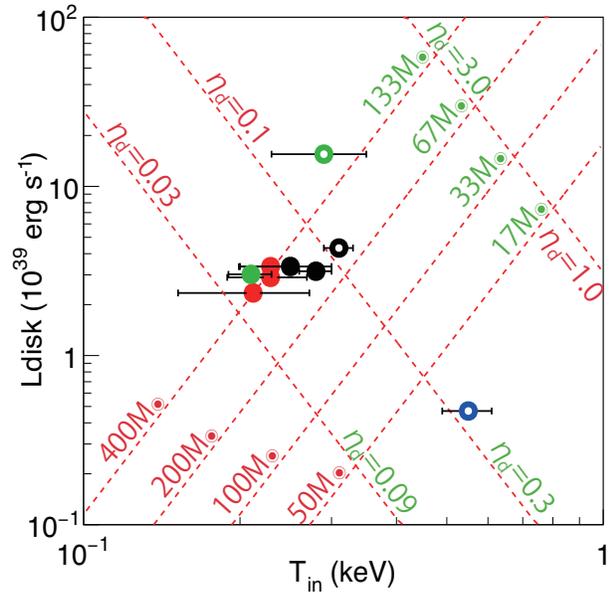}
\caption{The \textquotedblleft H-R diagram\textquotedblright \ of BHBs (Makishima et al. 2000) on which the disk luminosity (directly visible plus seed-photon source) of the present objects are plotted against their $T_{\rm{in}}$. The colors of data points are the same as in Fig. \ref{label2}. Red grids and numbers describe the mass and $\eta_{\rm{d}}$ (see text) when assuming Schwarzschild BHs with an accretion disks extending up to their ISCO, while green ones are when the disks are truncated at 3 times the ISCO.}
\label{label3}
\end{center}
\end{figure}
We can also estimate the masses of our sample ULXs from physics of the standard accretion disk by Shakura \& Sunyaev (1973). Figure \ref{label3} is the \textquotedblleft H-R diagram\textquotedblright \ invented by Makishima et al. (2000), where the analyzed spectra are plotted on the plane of $T_{\rm{in}}$ vs. the disk luminosity; the latter quantity is calculated by summing that of the directly visible disk, and that supplying the Compton seed photons. Dotted lines which go from bottom left to top right represent the Stefan-Boltzmann law
\begin{equation}
L_{\rm{disk}}=4\pi R_{\rm{in}}^{2}\sigma T_{\rm{in}}^{4}
\end{equation}
 for several BH masses. Here, $R_{\rm{in}}$ and $\sigma$ are the inner disk radius, and the Stefan-Boltzmann constant, respectively. Red grids are for non-rotating BHs of which $R_{\rm{in}}$ reaches 3 times the Schwarzschild radius $R_{\rm{S}}=2GM/c^{2}$, i.e. Innermost Stable Circular Orbit (ISCO). The other family of dashed lines which go from top left to bottom right represent the constant values of $\eta_{\rm{d}}$, namely, $L_{\rm{disk}}$ normalized to the Eddington luminosity.
 
 Although it is not yet clear at present whether the disk emission from ULXs (including our sample objects) behave as $L_{\rm disk} \propto T_{\rm in}^4$as predicted by the Stefan-Boltzmann law, let us assume as a working hypothesis that this approximately holds. Then, our sample objects except M33 X-8 are consistent with $200\--400~M_{\rm{\odot}}$ non-rotating BHs shining at $\eta_{\rm{d}}=0.03-0.1$. Since the total luminosity is typically $3\--10$ times higher than $L_{\rm disk}$, the overall Eddington ratio becomes $\eta = 0.3\--0.5$.
Therefore, they are inferred to be somewhat massive BHs of which the disk luminosity is sub-Eddington but the total luminosity is close to (within a factor of 3) the Eddington limit, 
including a possibility of mildly super-Eddington emission. M33 X-8 is inferred to be a $\sim 25~M_{\rm{\odot}}$ BH, slightly more massive than the ordinary stellar-mass BHs, radiating at $\eta_{\rm{d}}= 0.1$ and $\eta \sim 0.3$ which is in agreement with the expectation mentioned in \S 2.4.
 
 We have so far assumed $R_{\rm{in}}=$ISCO. However Tamura et al.~(2012) reported that one representative BHB called GX339-4 showed a disk truncation at 2.2 times the ISCO while it was in Very High State (VHS), which emerges in relatively high $\eta$ regime. If we assume that ULXs are in somewhat similar accretion mode to the VHS, we may allow the inner-disk radius to be truncated at $\sim 3$ times the ISCO. Then, the values of mass and $\eta_{\rm{d}}$ becomes as indicated by green labels in Fig. \ref{label3}. Here, we employed eq.(9) and eq.(11) of Makishima et al.~(2000). In this case, the sample ULXs are inferred to be still massive but shining at closer to the Eddington limit. This conclusion, which is based on our particular choice of the spectral modeling ($\tt{diskbb}+\tt{nthcomp}$), is fully consistent with that derived in \S4.2.1 in a more model-independent manner.   
%\newpage%%%%%%%%%%%%%%%%%%%%%%%%%%%%%%%%%%%%%%%%%%%%%%%%%%%%%%

\end{document}